\documentclass[aps,pra,showpacs,superscriptaddress,nofootinbib]{revtex4}
\usepackage{bbm}
\usepackage{mathrsfs}
\usepackage{epsfig,psfrag}
\usepackage{amsmath,amsfonts,amssymb}
\usepackage[usenames]{color}
\usepackage{bm}
\usepackage{ulem}
\def\mSe{ce}\def\mSo{se}
\def\mJe{Je}\def\mJo{Jo}

\def\mHe{He}\def\mHo{Ho}
\def\mIe{Ie}\def\mIo{Io}
\def\mKe{Ke}\def\mKo{Ko}

\begin{document}

\title{Electromagnetic Casimir Forces in Elliptic Cylinder Geometries}

\pacs{42.25.Fx, 03.70.+k, 12.20.-m}

\author{Noah Graham}
\email{ngraham@middlebury.edu}
\affiliation{Department of Physics,
Middlebury College,
Middlebury, VT 05753, USA}

\begin{abstract}
The scattering theory approach makes it possible to carry out exact
calculations of Casimir energies in any geometry for which the
scattering $T$-matrix and a partial wave expansion of the free Green's
function are available.  We implement this program for the case of a
perfectly conducting elliptic cylinder, thereby completing the set of
geometries where electromagnetic scattering is separable.
Particular emphasis is placed on the case of zero radius, where the
elliptic cylinder reduces to a strip.
\end{abstract}

\maketitle

\section{Introduction}

Formulating the Casimir energy in terms of scattering theory has made
it possible to efficiently reduce quantum field theory calculations to
standard problems in quantum mechanics and electromagnetism.
By expressing the ``TGTG'' form of the Casimir energy
\cite{Kenneth06} in appropriate scattering bases, one can calculate
the Casimir interaction energy of a collection of objects as a
combination of the objects' scattering amplitudes ($T$-matrices)
together with universal translation matrices, which are obtained from
a mode expansion of the free Green's function
\cite{spheres,scalar,universal}.  The former are computed for each object
individually, while the latter depend only on the objects'
relative positions and orientations.  As a result, the Casimir energy 
can be computed for any collection of objects for which the scattering
$T$-matrix is available within a standard scattering basis.  This
approach allows for exact calculations, extending earlier results
using asymptotic expansions \cite{Balian77,Balian78} and results from
scalar theories \cite{Bulgac01,Bulgac06,Wirzba08}.  It can also be
applied in the weak coupling approximation \cite{Milton08-1}.  For
objects without special symmetries, however, one must ultimately turn
to computational methods to compute either the $T$-matrix or the
associated Green's functions \cite{Johnson,Gies1,Gies2,Forrow:2012sp}.

With sufficient symmetry, the exact $T$-matrix can take an
analytically calculable form, greatly reducing the amount of
computation required.  This reduction has made it possible to apply
the scattering method to efficient computations of 
the Casimir energy for planes \cite{Lambrecht06},
spheres and ordinary cylinders \cite{Emig06,spheres,universal,Teo:2012kf},
parabolic cylinders \cite{parabolic1,parabolic2}, and wedges and cones
\cite{wedge}.  Here we complete the set of separable geometries in
electromagnetism by treating the case of an elliptic cylinder.  This
geometry has been investigated for microfabricated materials
using a Lifshitz formula approach in Ref.~\cite{Decca} and has been
used to study Casimir self-energies in Refs.~\cite{Kitson:2006hf,Straley}.

\section{Scattering in Elliptic Cylinder Coordinates}

We begin by formulating scattering theory in elliptic cylinder
coordinates,
\begin{eqnarray}
 x &=& d\cosh \mu\cos \theta \cr
 y &=& d\sinh \mu\sin \theta \,,
\end{eqnarray}
where $2d$ is the interfocal separation of our elliptic cylinder
coordinates, $\theta$ is the analog of the angle in ordinary
cylindrical coordinates, and $\mu$ is the analog of the radius, with
\begin{equation}
r = \sqrt{ x^2 +  y^2} = 
d\sqrt{\frac{\cosh 2 \mu+\cos 2 \theta}{2}}\to \frac{d}{2}
e^{ \mu} 
\end{equation}
as $ \mu\to\infty$.

We use separation of variables to form solutions of the Helmholtz
equation $-\nabla^2 \psi(\bm{r}) = k^2\psi(\bm{r})$
as products of functions of $\mu$, $\theta$ and $z$
individually.  For the functions of $z$, we have ordinary complex
exponentials $e^{ik_z z}$, which will multiply angular functions of
$\theta$ and radial functions of $\mu$.  Since we have parity
symmetry, we can choose our angular solutions to be either even or odd
under reflection across the $x$-axis, $\theta \to -\theta$.  Unlike the
ordinary cylinder case, the elliptic angular solutions depend on the
wave number $k$, and the elliptic radial solutions associated with the
the odd and even angular solutions differ and depend on the wave
number and radius separately, rather than only on the product $kr$.  For
$q=\frac{d^2}{4}(k^2-k_z^2)$, the angular solutions are the even
and odd angular Mathieu functions $\mSe_m(\theta, q)$ and
$\mSo_m(\theta, q)$, which are the analogs of $\cos m \theta$ and
$\sin m \theta$ respectively.  As in the case of ordinary cylindrical
coordinates, for the even functions $m$ runs from $0$ to $\infty$,
while for the odd functions $m$ runs from $1$ to $\infty$.  For the
corresponding radial functions, we have both the even and odd first
kind solutions $\mJe_m(\mu, q)$ and $\mJo_m(\mu, q)$, the analogs of
the Bessel function $J_m(\sqrt{k^2-k_z^2}r)$, and the even and odd
outgoing wave solutions $\mHe_m(\mu, q)$ and $\mHo_m(\mu, q)$, the
analogs of the Hankel function $H_m^{(1)}(\sqrt{k^2-k_z^2}r)$.  We will
normalize the Mathieu functions so that they obey the same
orthonormality conditions as their cylindrical analogs, except that
the $m=0$ even angular function will be normalized so that its root
mean square average value is $1/\sqrt{2}$ (the same as for all the
other angular functions) instead of $\cos 0 = 1$.  As a result, we have
\begin{equation}
\int_0^{2\pi} \mSe_m(\theta, q)^2 d\theta  = \int_0^{2\pi}
\mSo_m(\theta, q)^2 d\theta = \pi \,,
\end{equation}
with the radial functions normalized to coincide with their cylindrical
analogs asymptotically.  Our notation and normalization match that of
Ref.~\cite{Graham:2005cq}, which defines Mathieu functions following
the conventions of Abramowitz and Stegun \cite{Abramowitz}, but uses a 
modified notation that is more closely analogous to the ordinary
cylinder case.  We will make use of identities for elliptic cylinder
functions found in standard references \cite{Abramowitz,Morse53,Bateman}.

The key ingredients for our calculation will be the free Green's function
\begin{eqnarray}
G(\bm{r}_1, \bm{r}_2, k)
&=& \int_{-\infty}^\infty \frac{d k_z}{2 \pi}
\frac{i}{2} \left[
\sum_{m=0}^\infty 
\mSe_m(\theta_1, q) \mSe_m(\theta_2, q) \mJe_m(\mu_<, q) \mHe_m(\mu_>, q)
\right. \cr && \left. + \sum_{m=1}^\infty 
\mSo_m(\theta_1, q) \mSo_m(\theta_2, q) \mJo_m(\mu_<, q) \mHo_m(\mu_>, q)
\right] \,,
\label{eqn:Green1}
\end{eqnarray}
where $ \mu_<$ ($ \mu_>$) is the smaller (larger) of
$ \mu_1$ and $ \mu_2$, and the expansion of a plane wave,
\begin{equation}
e^{i \bm{k} \cdot {\bm{r}}} = e^{ik_z  z} \left[
2 \sum_{m=0}^\infty 
i^m \mSe_m(\phi, q) \mSe_m(\theta, q) \mJe_m(\mu, q) 
+ 2 \sum_{m=1}^\infty 
i^m \mSo_m(\phi, q) \mSo_m(\theta, q) \mJo_m(\mu, q) 
\right]\,,
\label{eqn:plane1}
\end{equation}
where $\mu$, $ \theta$, and $z$ are the elliptic cylinder
coordinates of ${\bm{r}}$ and $\phi = \arctan \frac{k_y}{k_x}$
is the angle of $\bm{k} = (k_x,k_y,k_z)$ in the $xy$-plane, with
$k^2 = k_x^2 + k_y^2 + k_z^2$.

We will work on the imaginary $k$-axis $k=i\kappa$, so that $k_y =
i\sqrt{\kappa^2+k_x^2+k_z^2}$ and  $q = -d^2(\kappa^2 + k_z^2)/4$ 
is negative.  As a result, it is convenient to rewrite these
expressions in terms of modified radial functions,
\begin{eqnarray}
G(\bm{r}_1, \bm{r}_2, k)
&=& \int_{-\infty}^\infty \frac{d k_z}{2 \pi}
\frac{1}{\pi} \left[
\sum_{m=0}^\infty 
\mSe_m(\theta_1, q) \mSe_m(\theta_2, q) \mIe_m(\mu_<, -q) \mKe_m(\mu_>, -q)
\right. \cr && \left. + \sum_{m=1}^\infty 
\mSo_m(\theta_1, q) \mSo_m(\theta_2, q) \mIo_m(\mu_<, -q) \mKo_m(\mu_>, -q)
\right]
\label{eqn:Green2}
\end{eqnarray}
and
\begin{equation}
e^{i \bm{k} \cdot {\bm{r}}} = e^{ik_z  z} \left[
2 \sum_{m=0}^\infty 
(-1)^m \mSe_m(\phi, q) \mSe_m(\theta, q) \mIe_m(\mu, -q) 
+ 2 \sum_{m=1}^\infty 
(-1)^m \mSo_m(\phi, q) \mSo_m(\theta, q) \mIo_m(\mu, -q) 
\right]\,,
\label{eqn:plane2}
\end{equation}
where $\mIe_m(\mu, -q) = i^{-m} \mJe_m(\mu, q)$,
$\mIo_m(\mu, -q) = i^{-m} \mJo_m(\mu, q)$,
$\mKe_m(\mu, -q) = i^{m+1} \frac{\pi}{2} \mHe_m(\mu, q)$ 
and $\mKo_m(\mu, -q) = i^{m+1} \frac{\pi}{2} \mHo_m(\mu, q)$ 
are the modified outgoing radial functions.  

We will consider scattering with Dirichlet and Neumann boundary
conditions on an elliptic cylinder of radius $\mu_0$.  For the
scattering amplitudes, we have
$\displaystyle {\cal T}_{m k_z m' k_z'}^{e,o} = 
2\pi \delta(k_z - k_z')\delta_{mm'} {\cal T}_m^{e,o}$, with
\begin{eqnarray}
{\cal T}_m^e =
-\frac{\mIe_m \left(\mu_0, -q \right)} {\mKe_m\left(\mu_0, -q \right)} \qquad
{\cal T}_m^o =
-\frac{\mIo_m \left(\mu_0, -q \right)} {\mKo_m\left(\mu_0, -q \right)}
&& \hbox{\qquad (Dirichlet)} \cr
{\cal T}_m^e =
-\frac{\mIe_m' \left(\mu_0, -q \right)} {\mKe_m' \left(\mu_0, -q \right)} \qquad
{\cal T}_m^o =
-\frac{\mIo_m' \left(\mu_0, -q \right)} {\mKo_m' \left(\mu_0, -q \right)}
&& \hbox{\qquad (Neumann),}
\end{eqnarray}
where prime indicates a derivative with respect to $\mu$.

\section{Elliptic Cylinder and Plane}

To consider the elliptic cylinder's interaction with a plane,
we will need to connect the elliptic cylinder and planar geometries.
To do so, we make use of the expression for the free Green's function
in Cartesian coordinates for $y_2>y_1$,
\begin{equation}
G(\bm{r}_1,\bm{r}_2, k) 
= \int_{-\infty}^\infty \frac{d k_z}{2 \pi} e^{i k_z(z_2 - z_1)}
\frac{i}{4\pi} \int_{-\infty}^\infty \frac{dk_x}{k_y} 
e^{i(k_x (x_2-x_1) + k_y (y_2 - y_1))} \,,
\label{eqn:Greenplane}
\end{equation}
where $k_y = \sqrt{k^2 - k_x^2 - k_z^2} = i \sqrt{\kappa^2 + k_x^2 +
k_z^2}$.  We equate Eq.~(\ref{eqn:Greenplane}) to 
the Green's function in Eq.~(\ref{eqn:Green2}), expand the plane wave 
$\displaystyle e^{i \bm{k}\cdot{\bm{r}_2}}$ in Eq.~(\ref{eqn:Greenplane})
using Eq.~(\ref{eqn:plane2}), make the substitution $k_x \to -k_x$,
and finally use the orthogonality of the
regular elliptic cylinder solutions to equate both sides term by term
in the sums over $m$.  The result is an expansion for the elliptic
outgoing wave solutions in terms of plane waves for $y<0$ \cite{Abramowitz},
\begin{eqnarray}
\mSe_m(\theta, q) \mKe_m(\mu, -q)
e^{ik_z z} &=& \int_{-\infty}^{\infty} dk_x \left[
\frac{i}{2 k_y} \mSe_m(\phi, q)\right]
e^{-i k_y y + i k_x x} e^{ik_z z}  \cr
\mSo_m(\theta, q) \mKo_m(\mu, -q)
e^{ik_z z} &=& \int_{-\infty}^{\infty} dk_x \left[
\frac{-i}{2k_y} \mSo_m(\phi, q)\right]
e^{-i k_y y + i k_x x} e^{ik_z z} \,.
\label{eqn:expandout}
\end{eqnarray}
The quantities in
brackets represent the translation matrix elements,
which we must then multiply by the normalization factor
$\frac{C^{\hbox{\tiny elliptic}}_m}{C^{\hbox{\tiny plane}}_{k_x}}$, 
where we can read off
$C^{\hbox{\tiny elliptic}}_m = \sqrt{\frac{1}{\pi}}$
and 
$C^{\hbox{\tiny plane}}_{k_x} = \sqrt{\frac{i}{4\pi k_y}}$
from the expressions for the free Green's function in 
Eqs.~(\ref{eqn:Green2}) and (\ref{eqn:Greenplane}).

Finally, the $T$-matrix elements for the plane in Cartesian
coordinates are simply ${\cal T}^P = \pm 1$ for Neumann and
Dirichlet boundary conditions respectively.  (For more general
boundary conditions on the plane, this scattering amplitude would be a
function of $k_x$.)

\begin{figure}[htbp]
\includegraphics[width=0.6\linewidth]{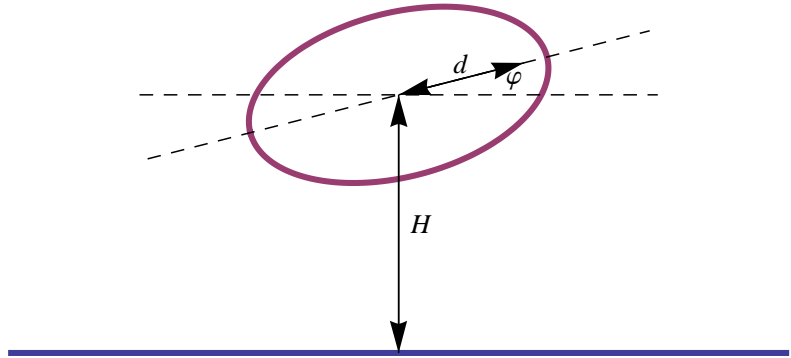}
\caption{Geometry for the elliptic cylinder and plane.}
\label{fig:tilt}
\end{figure}

We have now obtained the $T$-matrix elements, which describe how waves
scatter off each object individually, and the translation matrix
elements, which convert the scattering bases between the two objects.
As a result, we are prepared to assemble these ingredients into the
result for the full Casimir interaction energy per unit length.  We
consider a perfectly conducting plane oriented perpendicular to the
$y$-axis and a perfectly conducting elliptic cylinder
with its $z$-axis parallel to the plane, its center at a distance
$H$ from the plane, and its major axis at an angle $\varphi$ to the
plane, as shown in Fig.~\ref{fig:tilt}.  This angle represents a
rotation of the elliptic cylinder coordinates $\theta$ and $\mu$
relative to the Cartesian coordinates $x$ and $y$, which we then
implement in Eq.~(\ref{eqn:expandout}) by adding a constant shift
$\varphi$ to the angle $\phi=\arctan \frac{k_y}{k_x}$ in the
translation matrix elements.

For a particular choice of boundary conditions, we can now use the
approach of Refs.~\cite{spheres,scalar,universal} to write the Casimir
energy per unit length as
\begin{equation}
\frac{\cal E}{\hbar c L}=
\int_0^\infty \frac {d\kappa}{2 \pi} 
\int_{-\infty}^\infty \frac {dk_z}{2 \pi}
\log \det \left(\mathbbm{1}_{mm'}^{\chi \chi'} - 
{\cal T}_{m}^{\chi}
\int \frac{i d k_x}{k_y} 
{\cal U}_{m k_x}^{\chi}
{\cal T}^{P}_{k_x}
\hat {\cal U}_{m' k_x}^{\chi'}
\right)\,,
\end{equation}
where the matrix determinant runs over $\chi,\chi'=o,e$
with $m=0,1,2,3\ldots$ for $\chi=e$ and $m=1,2,3\ldots$ for $\chi=o$, and
similarly for $m'$ and $\chi'$.  The translation matrices ${\cal
U}^\chi_{m k_x}$ and reverse translation matrices $\hat {\cal
U}^\chi_{m k_x}$ are given by
\begin{equation}
{\cal U}^e_{m k_x} =
\mSe_m\left(\phi + \varphi, q\right) e^{ik_y H}
\qquad
\hat{\cal U}^e_{m k_x} =
\mSe_m\left(-\phi + \varphi, q\right) e^{ik_y H} 
\label{eqn:trans1}
\end{equation}
for the even modes and
\begin{equation}
{\cal U}^o_{m k_x} =
\mSo_m\left(\phi + \varphi, q\right) e^{ik_y H}
\qquad
\hat{\cal U}^o_{m k_x} =
\mSo_m\left(-\phi + \varphi, q\right) e^{ik_y H} 
\label{eqn:trans2}
\end{equation}
for the odd modes.

We can then change the integration variable from $k_x$ to
$u=\frac{1}{i} \left(\phi -\frac{\pi}{2}\right)$ and combine the
$\kappa$ and $k_z$ equations into a single integral over
$p=\sqrt{\kappa^2 + k_z^2}$, so that $q=-\frac{d^2 p^2}{4}$.  We
obtain
\begin{equation}
\frac{\cal E}{\hbar c L}=
\frac{1}{4\pi}\int_0^\infty p dp
\log \det \left[\mathbbm{1}_{mm'}^{\chi \chi'} - 
{\cal T}^\chi_{m} {\cal T}^P
\int_{-\infty}^\infty du e^{-2 p H \cosh u}
\genfrac{}{}{0pt}{}{\mSe_m}{\mSo_m}
\left(\frac{\pi}{2} + iu + \varphi, q\right)
\genfrac{}{}{0pt}{}{\mSe_{m'}}{\mSo_{m'}}
\left(\frac{\pi}{2} - iu + \varphi, q\right)
\right]\,,
\label{eqn:energy}
\end{equation}
where we choose $\mSe_m$ for $\chi=e$ and $\mSo_m$ for $\chi=o$, and
similarly for $m'$ and $\chi'$.  The full electromagnetic Casimir
energy is the sum of this result for Dirichlet
conditions on both surfaces and for Neumann
conditions on both surfaces.  Note that the established
result for an ordinary cylinder \cite{Emig06} can be obtained from
this expression by replacing the elliptic functions with their
ordinary cylindrical analogs, combining the even and odd modes
using $\cosh m u \cosh m' u + \sinh m u \sinh m' u  = \cosh (m+m')u$,
and employing the integral identity
\begin{equation}
K_{n}(\sigma)  = \int_0^\infty e^{-\sigma \cosh u} \cosh n u \, du \,.
\end{equation}

There are several special cases of interest in which the calculation
simplifies:
\begin{itemize}
\item Plane perpendicular to the ellipse's major axis.  

For $\varphi = \pi/2$, the elliptic cylinder's major axis runs
perpendicular to the plane.  By the reflection symmetry across the
$y$-axis, the even and odd sectors decouple, and we can compute the
Casimir energy by considering the odd and even elliptic modes
separately.

\item Plane parallel to the ellipse's major axis.

For $\varphi=0$, the elliptic cylinder's major axis lies parallel to the
plane.  This case also has reflection symmetry across the $y$-axis,
but this symmetry does not correspond directly to the symmetry of the
even and odd Mathieu functions.  Instead, the even Mathieu functions
of even order and the odd Mathieu functions of odd order are symmetric
under this transformation, while the odd Mathieu functions of even
order and the even Mathieu functions of odd order are antisymmetric.
(This is the same symmetry structure as the ordinary trigonometric
functions have when their argument is displaced by $\pi/2$.)
Thus we can again decompose the problem into two independent sectors,
consisting of the modes for which the parity of the elliptic functions
matches the parity of $m$, and the modes for which they are opposite.

\item Zero radius cylinder.

An elliptic cylinder with $\mu_0=0$ becomes a strip of width $2d$,
allowing us to study the effects of edges
\cite{wedge,parabolic1,parabolic2,Gies3,Kabat1}.
In that case we have ${\cal T}_m^o =0$ for a Dirichlet boundary and 
${\cal T}_m^e =0$ for a Neumann boundary, since in these cases the
free modes already obey the boundary condition at the surface.  These
modes therefore give zero contribution to the Casimir energy in this
case, and can be omitted from the calculation.

\end{itemize}

\section{Numerical Results}

We can now compute the Casimir energy by straightforward numerical
integration of Eq.~(\ref{eqn:energy}).  To compute the modified radial
functions needed for the scattering amplitude, we use the package of
Alhargan \cite{Alhargan:2000,Alhargan:2000a}.  (Standard packages such
as Maple and Mathematica only implement the angular Mathieu functions
of the first kind.  Although the radial functions are related to the
angular functions with imaginary argument, without an implementation
of the second kind angular function we cannot take advantage of this
relationship to compute the functions needed to for the scattering
amplitude.)  The angular functions arising from the translation
matrix, on the other hand, need to be computed for complex arguments,
which are not supported directly in the Alhargan package.
Fortunately, since only the first kind angular functions are required,
we can use the implementation in Mathematica, which supports fully
complex arguments.  As a final complication, because of problems with
the Mathieu function routines in the current version of Mathematica
for the case where the parameter $q<0$, we make use of the identities
\begin{eqnarray}
\mSe_m(\mu,q) &=& 
\left\{ \begin{array}{l@{\quad}l}
(-1)^{\frac{m}{2}} \mSe_m\left(\frac{\pi}{2} - \mu, -q\right) &
\hbox{for $m$ even} \cr
(-1)^{\frac{m-1}{2}} \mSo_m\left(\frac{\pi}{2} - \mu, -q\right) &
 \hbox{for $m$ odd}
\end{array} \right. \cr
\mSo_m(\mu, q) &=& 
\left\{ \begin{array}{l@{\quad}l}
(-1)^{\frac{m}{2}-1} \mSo_m\left(\frac{\pi}{2} - \mu, -q\right) &
\hbox{for $m$ even} \cr
(-1)^{\frac{m-1}{2}} \mSe_m\left(\frac{\pi}{2} - \mu, -q\right) &
 \hbox{for $m$ odd}
\end{array} \right.
\end{eqnarray}
so that we only need to compute the angular functions for $-q>0$.
As a result, after importing the Alhargan routines for the modified
radial functions, it is possible to carry out the full calculation within
Mathematica.  Because of limitations in the ability of the angular
routines to handle large imaginary arguments, however, it was not
possible to extend the calculation to very small separations.

\begin{figure}
\includegraphics[width=0.5\linewidth]{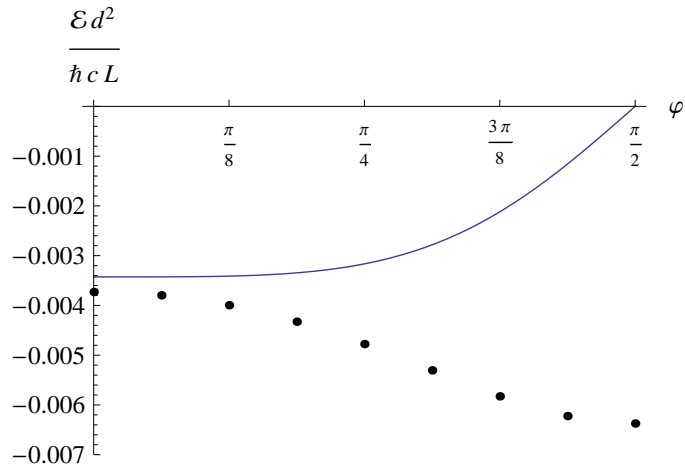}
\caption{Electromagnetic Casimir interaction energy for a perfectly
conducting strip opposite a perfectly conducting plane, as a function
of the orientation angle $\varphi$.  The distance $H$ from the
center of the strip to the plane is twice the distance from the center
of the strip to the edge of the strip, $H=2d$.  The solid
line shows the proximity force approximation.}
\label{fig:rotate}
\end{figure}

Figure~\ref{fig:rotate} shows the orientation dependence of
the Casimir interaction energy for a perfectly conducting strip (an
elliptic cylinder of zero radius) for the case where the distance $H$
from the center of the strip to the plane is twice the distance from
the center of the strip to the edge of the strip, $H=2d$.  Because
higher values did not change the results appreciably, the matrix
determinants were truncated at $m_{\hbox{\tiny max}} = 8$.  We see that
the lowest energy occurs for $\varphi=\pi/2$, when the strip is
perpendicular to the plane.  As expected, the result for the energy
per unit length in this case, $\frac{{\cal E} d^2}{\hbar c L}
= -0.00637$, is less negative than the $-0.00674$
one finds \cite{parabolic1,parabolic2} for the case where the strip is
extended to an infinite half-plane whose edge maintains the same
distance $H-d=d$ from the infinite plane.  We note, however, that if we
subtract the contribution from a half-plane at distance $H+d=3d$ from the
result for the half-plane at distance $H-d=d$ to account for the
missing remainder of the half plane, we obtain $-0.00674
\cdot \frac{8}{9} = -0.00599$, which underestimates the magnitude
of the true result for the strip.  We also compare these results 
to the proximity force approximation (PFA),
\begin{equation}
\frac{{\cal E}_{PFA}^{(0)}}{\hbar c L}=
-\frac{\pi^2}{720} \int_{-d\cos\varphi}^{-d\cos\varphi}
\frac{dx}{(H+x \tan \varphi)^3} =
-\frac{\pi^2}{360} \frac{Hd \cos \varphi}{\left(
H^2 - d^2 \sin^2 \varphi\right)^2}
\end{equation}
which gives a good approximation for $\varphi=0$ but goes to zero at 
$\varphi=\pi/2$.  For $\varphi\neq0$ the derivative expansion
correction to the PFA \cite{Fosco:2011xx,beyondpfa} is also invalid,
because of the sharp curvature at the point of closest approach.

\begin{figure}
\includegraphics[width=0.5\linewidth]{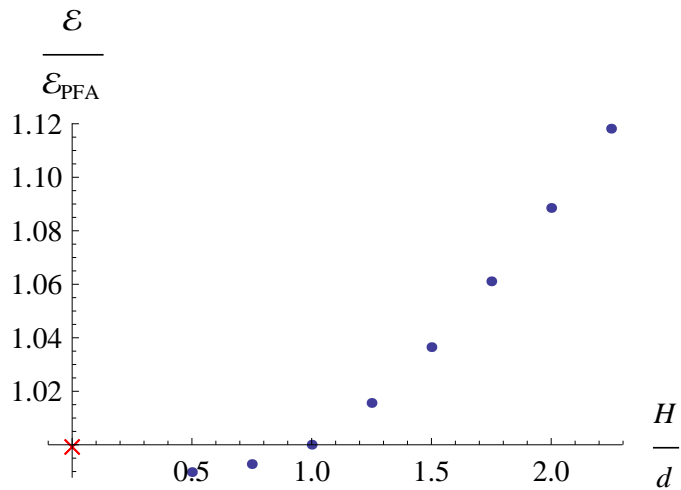}
\caption{Ratio of the electromagnetic Casimir interaction energy to
the proximity force approximation (PFA) for a perfectly
conducting strip of width $2d$ parallel to a perfectly conducting
plane, as a function of the separation $H$.  As in the case of an
ordinary cylinder \cite{Emig06}, the ratio is a nonmonotonic function
of $H$.}
\label{fig:pfa}
\end{figure}

Figure \ref{fig:pfa} shows the Casimir interaction energy for a strip
oriented parallel to a plane as a function of the
distance to the plane.  The energy is shown as a ratio with the PFA
result (in this case the correction from the derivative expansion
vanishes).  As in the case of the ordinary cylinder \cite{Emig06}, the
PFA is an underestimate at large distances, but at short distances the
exact result approaches the PFA result from below.  These
calculations were carried out with the matrices truncated at several
different values of $m_{\hbox{\tiny max}}$ up to $m_{\hbox{\tiny max}}
= 16$, with the final result then obtained by extrapolating these
results for $m_{\hbox{\tiny max}} \to \infty$.

\section{Discussion}

We have computed the Casimir interaction energy for an elliptic
cylinder, the last remaining geometry for which electromagnetic
scattering is separable.  For a plane, cylinder, and sphere,
the problem remains separable even for a dielectric, while for a
parabolic cylinder, elliptic cylinder, wedge, and cone only perfect
conductors can be solved exactly.  However, the scattering method is
particularly useful in these latter cases, because they contain sharp
limits in which the PFA is invalid. In principle, it should be
possible to extend the elliptic cylinder result to a hyperbolic
cylinder in the same way as the wedge is obtained from the ordinary
cylinder and the cone is obtained from the sphere, but at present
there do not appear to be routines available for computing all the
Mathieu functions of complex order that would be needed for such a
calculation.

Focusing on the limit in which the elliptic cylinder becomes a
strip has made it possible to study the orientation dependence of the
Casimir force, to show how the PFA depends on distance and angle, and 
to observe non-superposition effects in the perpendicular
configuration (where the PFA is invalid).  With improvements to the
available routines for computing Mathieu functions, this calculation
could offer an independent check of the edge correction that was
obtained for half-planes \cite{parabolic1,parabolic2,planes}.  More
generally, this calculation establishes another addition to the
toolbox of Casimir problems that can be cast into analytically
tractable form.
\section{Acknowledgements}

N.\ G. thanks T.\ Emig, R.\ L.\ Jaffe, M.\ Kardar, and K.\ Milton
for helpful conversations and suggestions.  This work was supported in
part by the National Science Foundation (NSF) through grant PHY-1213456.

\bibliographystyle{apsrev}
\bibliography{article}

\end{document}